\documentclass[dvips,aps,prl,reprint,showpacs,groupedaddress,superscriptaddress]{revtex4-1} 

\usepackage{multirow}
\usepackage{amsmath,amssymb}
\usepackage[dvips]{graphicx}

\begin{document} 
 
\title{High-precision determination of the electric and magnetic form  
  factors of the proton} 
  
 \author{J.\,C. Bernauer}\email{bernauer@mit.edu}  
\altaffiliation{Present address: Laboratory for Nuclear Science, MIT, Cambridge,  MA 02139, USA.}

\author{P.~Achenbach}
\author{C.~{Ayerbe Gayoso}}
\author{R.~B\"ohm}\affiliation{Institut f\"ur Kernphysik, Johannes Gutenberg-Universit\"at Mainz, 55099 Mainz, Germany.}
\author{D.~Bosnar}\affiliation{Department of Physics, University of Zagreb, 10002 Zagreb, Croatia.}  
\author{L.~Debenjak}\affiliation{Jo\v zef Stefan Institute, Ljubljana, Slovenia.}  
\author{M.\,O.~Distler}\email{distler@kph.uni-mainz.de}
\author{L.~Doria}
\author{A.~Esser}\affiliation{Institut f\"ur Kernphysik, Johannes Gutenberg-Universit\"at Mainz, 55099 Mainz, Germany.}  
\author{H.~Fonvieille}\affiliation{LPC-Clermont, Universit\'e Blaise
  Pascal, CNRS/IN2P3, F-63177 Aubi\`{e}re Cedex, France.}  
\author{J.\,M.~Friedrich}\affiliation{Physik-Department, Technische Universit\"at M\"unchen, 85748 Garching, Germany}  
\author{J.~Friedrich}
\author{M.~{G\'omez Rodr\'iguez de la Paz}}\affiliation{Institut f\"ur Kernphysik, Johannes Gutenberg-Universit\"at Mainz, 
            55099 Mainz, Germany.}
\author{M.~Makek}\affiliation{Department of Physics, University of Zagreb, 10002 Zagreb, Croatia.}  
\author{H.~Merkel}
\author{D.\,G.~Middleton}
\author{U.~M\"uller}
\author{L.~Nungesser}
\author{J.~Pochodzalla}\affiliation{Institut f\"ur Kernphysik, Johannes Gutenberg-Universit\"at Mainz, 55099 Mainz, Germany.}  
\author{M.~Potokar}\affiliation{Jo\v zef Stefan Institute, Ljubljana, Slovenia.}  
\author{S.~{S\'anchez Majos}}
\author{B.\,S.~Schlimme}\affiliation{Institut f\"ur Kernphysik, Johannes Gutenberg-Universit\"at Mainz, 55099 Mainz, Germany.}  
\author{S.~\v{S}irca} \affiliation{Department of Physics, University of Ljubljana, Slovenia.}
\affiliation{Jo\v zef Stefan Institute, Ljubljana, Slovenia.}  
\author{Th.~Walcher}
\author{M.~Weinriefer}\affiliation{Institut f\"ur Kernphysik, Johannes Gutenberg-Universit\"at Mainz, 55099 Mainz, Germany.}  
\collaboration{A1 Collaboration}\noaffiliation

\date{July 28, 2010}

\begin{abstract} 
New precise results of a measurement of the elastic electron-proton scattering cross  
section performed at the Mainz Microtron MAMI are presented. About 1400  
cross sections were measured with negative four-momentum transfers squared  
up to $Q^2=1\,(\mathrm{GeV}/c)^2$ with statistical errors below 0.2\%.  
The electric and magnetic form factors of the proton were extracted  
by fits of a large variety of form factor models directly to the cross  
sections. The form factors show some features at the scale of the  
pion cloud. The charge and magnetic radii are determined to be  
$\left\langle r_{E}^2\right\rangle^\frac{1}{2}=0.879(5)_\mathrm{stat.}(4)_\mathrm{syst.}
(2)_\mathrm{model}(4)_\mathrm{group}\,\mathrm{fm}$ and 
$\left\langle r_{M}^2\right\rangle^\frac{1}{2}=0.777 
(13)_\mathrm{stat.}(9)_\mathrm{syst.}(5)_\mathrm{model}(2)_\mathrm{group}\,\mathrm{fm}$. 
\end{abstract}

 
\maketitle  
 
The proton is composed of relativistic light constituents, 
i.e.\, quarks and gluons, as manifest in ``deeply inelastic'' lepton
scattering experiments. On the other hand, its global charge and
magnetization distributions are contained in the form factors of
elastic electron scattering. These provide a very instructive
connection between models based on effective degrees of freedom and the 
QCD-based parton theory of the nucleon \cite{Vanderhaeghen:2010nd}.  A particularly 
intriguing question is the existence of a direct signal for a pion cloud 
in the form factor hypothesized on the basis of pre-2003 data in
Ref.\,\cite{fw03}. Since Yukawa the virtual pion is considered as the
mediator of the nucleon-nucleon force. Today it is viewed as an
effective degree of freedom of the nucleon originating from the
spontaneous breaking of chiral symmetry producing the Godstone bosons
of QCD. An exciting question, addressed in this letter, is
whether it could be seen directly.
 
A quantity particularly sensitive to the existence of a pion cloud is
the proton charge radius. The determinations from electronic
hydrogen Lamb shift measurements \cite{Melnikov,Udem97} agree
with those from e-p scattering if Coulomb and
relativistic corrections are applied
\cite{Rosenfelder99,Sick03,Blunden}. This work presents a novel
determination of the electric and magnetic form factors and radii
using a direct fit of form factor models to the electron scattering
cross sections.
 
The Mainz accelerator MAMI provides a cw electron beam with energies up to  
1600\,MeV with very small halo and good energy definition. However, in this  
experiment only electron energies up to 855\,MeV were used since the higher 
energies were not yet available at the time of the experiment. Taking 
advantage of the three high-resolution spectrometers of the A1 collaboration 
it was possible to measure the elastic electron-proton scattering cross
section  with a statistical precision of better than 0.2\% and
extract the form factors up to a negative four-momentum transfer
squared of $Q^2=0.6\,(\mathrm{GeV}/c)^2$.
 
About 1400 cross sections were measured at beam energies of 180,  
315, 450, 585, 720, and 855\,MeV covering $Q^2$ from 0.004 to 
1\,$(\mathrm{GeV}/c)^2$. In order to achieve high accuracy the experiment 
aimed to maximize the redundancy of the data. For the angular scans the spectro\-meter 
angles were varied only in small steps so that the same scattering angle is 
measured up to 4 times with different regions of the spectrometer
acceptance, and parts of the angular range were measured with two spectrometers. 
The broad range of beam currents from below 1\,nA to more than 10\,$\mu$A, 
a consequence of the large range of cross sections, required special attention to 
the determination of the luminosity. Therefore, the current was measured  
redundantly with a fluxgate magnetometer and with a pA-meter connected to 
a collimator just downstream the electron source. Furthermore, the relative 
luminosity was measured at all times with one of the three spectrometers 
at a fixed scattering angle. 
 
In lowest order, the elastic electron-proton scattering cross section is 
described by the Rosenbluth formula 
\begin{equation} 
\left(\frac{\mathrm{d}\sigma}{\mathrm{d}\Omega}\right) =
\left(\frac{\mathrm{d}\sigma}{\mathrm{d}\Omega}\right)_\mathrm{Mott}
 \frac{\varepsilon G_E^2+\tau G_M^2}{\varepsilon\left(1+\tau\right)} 
\label{eqrosen}, 
\end{equation} 
where $G_E$ and $G_M$ are the electric and magnetic Sachs form factors, 
$m_p$ is the proton mass, $\tau=Q^2/(4m_p^2c^2)$ and  
$\varepsilon=\left(1+2\left(1+\tau\right)\tan^2(\theta/2)\right)^{-1}$ 
with the electron scattering angle $\theta$. 
However, also electromagnetic processes of higher order contribute to
the measured cross section, such as multiple photon exchange, vacuum
polarization, vertex corrections, and the radiation of a real photon
from the electron (Bethe-Heitler) or the proton (Born).
 
The code simulating the cross-section integration over the acceptance  
includes these processes following the description of 
ref.\,\cite{Vanderhaeghen2000} which gives results compatible with 
ref.\,\cite{Maximon2000}. Our approach extends this by an explicit  
calculation of the Feynman graphs of the Bethe-Heitler and Born processes  
on the event level. The simulation uses the standard dipole parametrization 
\begin{equation} 
G_E=\frac{G_M}{\mu_p}=G_\mathrm{std.\ dip.}=\left(1+\frac{Q^2}{0.71 
  (\mathrm{GeV}/c)^2}\right)^{-2} 
\end{equation} 
as a sufficient approximation for the true form factors ($\mu_p$ is
the proton's magnetic moment divided by the nuclear magneton). 
The division of the measured number of elastically scattered electrons
by luminosity  and simulated acceptance-integrated (standard dipole)  
cross-section yields the measured normalized cross section; this 
procedure accounts for the radiative processes.
 
Furthermore, Coulomb corrections according to ref.\,\cite{tsai61} have been 
applied, but no correction for the exchange of two hard photons (TPE)  
in the scattering process since no unique prescription exists yet.
 
The energy of the elastically scattered electrons $E_\mathrm{out}$ is  
shifted by the recoil energy of the proton.  The finite  
resolution of the spectrometers, the external energy-loss processes,  
and the internal bremsstrahlung widen the peak in the $\Delta E =  
E_\mathrm{out, calc.} - E_\mathrm{out, meas.}$ spectrum. 
The elastic events are selected by a cut around this peak. 
 
\begin{figure} 
\includegraphics[]{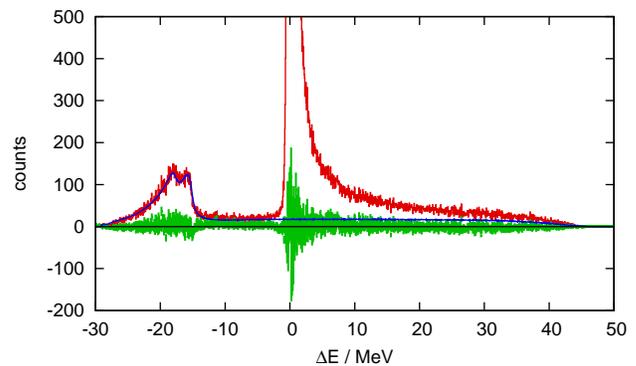} 
\caption{\label{fig_back} Typical $\Delta E$ spectrum. Red histogram: 
  measured spectra (height of elastic peak: 5500), blue line:  
  simulated background, green shaded band: data minus simulated  
  hydrogen peak and simulated background. The width of the band denotes the
  statistical uncertainty of the data. Data were measured at $E_0=450$\,MeV, 
  $Q^2=0.045(\mathrm{GeV}/c)^2$.} 
\end{figure}

The targets used were 2 and 5\,cm long cells filled with liquid
hydrogen with walls made of 10\,$\mu$m thick havar foil. The
primary source of background is the elastic and quasielastic
scattering off the nuclei in these walls. As verified with
empty-cell measurements, inelastic peaks are either small or outside
the cut region. The amplitude of its simulated shape was fitted
to the measured $\Delta E$ spectrum together with the simulated
hydrogen peak. As shown in Fig.\,\ref{fig_back} the simulated shapes
describe the measured spectra very well. Therefore, after subtraction
of the background from the raw spectra, the result is very insensitive 
to a variation of the cut around the elastic peak. In particular it shows 
the validity of the radiative corrections. The background contribution 
reaches up to 10\% but is below 4\% for most of the data.
 
For fixed $Q^2$ Eq.\,(\ref{eqrosen}) can be rewritten as an equation  
of a straight line in $\varepsilon$ in which $G_E^2$ and $G_M^2$ are  
fit parameters. This ``Rosenbluth''  method is model independent to  
first order in the photon propagator. However, one has to take the data at
constant $Q^2$ for a sufficiently broad range of $\varepsilon$. For a given 
energy and angular range this limits severely the kinematical range covered.  
This restriction can be avoided by a direct least-squares fit of models for 
$G_E$ and $G_M$ to the measured cross sections for all $Q^2$ and $\theta$. 
Of course, it is mandatory to select a wide range of different models and 
test for any model dependence. In this work we have used single-dipole and  
sum-of-two-dipoles models, the phenomenological parametrization of  
ref.\,\cite{fw03}, several variations of polynomial, reciprocal of polynomial,  
and spline based models, and a variant of the extended Gari-Kr\"umpelmann  
model \cite{Lomon01}.  A detailed study of the model dependence and  
the analysis method was performed using pseudo data generated according  
to a variety of previous parametrizations \cite{Arrington03,Arrington07,fw03}.  
This study showed that all models with enough flexibility, i.e.\, polynomials,  
their reciprocals, and splines, are able to reproduce the input with an error  
smaller than the other systematic errors. 
 
Since a determination of the absolute normalization of the measurement to 
better than 1\% is not possible, the normalizations of the individual
cross-section data sets are left as free parameters with the constraint 
$G_{E(M)}(Q^2=0)=1(\mu_p)$. The fitted normalizations are well within
the estimated uncertainty of 4\% and have almost no dependency on 
the model. For the flexible models the spread between the largest and smallest 
normalization factor is below 0.3\%. 
 
The results of the direct fits are compatible with the results from a 
classic Rosenbluth separation where such a comparison is possible.  
However, we find that the Rosenbluth approach is more sensitive  
to systematic deviations and is therefore a less robust estimator  
of $G_E^2$ and $G_M^2$. However, the Rosenbluth separation allows us to  
check for deviations from a straight line caused by possible problems  
in the data or by higher-order processes like TPE. At the level of  
the uncertainty of the measurements no systematic deviations  
from straight lines were found. 
 
As the direct fits of models are nonlinear, standard error estimation  
techniques for the fit are not guaranteed to be exact. Therefore, the  
confidence bands were calculated with the Monte Carlo technique  
including the errors of the normalizations. We find that Monte Carlo  
and the linearization used in standard error propagation yield almost  
identical results for all but one model. The confidence bands presented 
here are the widely used pointwise bands, meaning that one expects the 
true curve to be with 68\% probability within the band at any given single 
$Q^2$, but not necessarily at all $Q^2$ simultaneously. The Monte Carlo 
approach also allows one to construct simultaneous bands meaning that 
with 68\% probability the true curve does not leave the band for the full 
range of $Q^2$. It is somewhat involved to treat this problem with standard 
analytical methods \cite{Bernauer2010:ab}. The simultaneous  
bands can be obtained from the pointwise bands shown here by scaling  
the latter by a factor of around 2.3 for the $Q^2$ range up to  
0.6\,$(\mathrm{GeV}/c)^2$. 
 
\begin{figure} 
\includegraphics[]{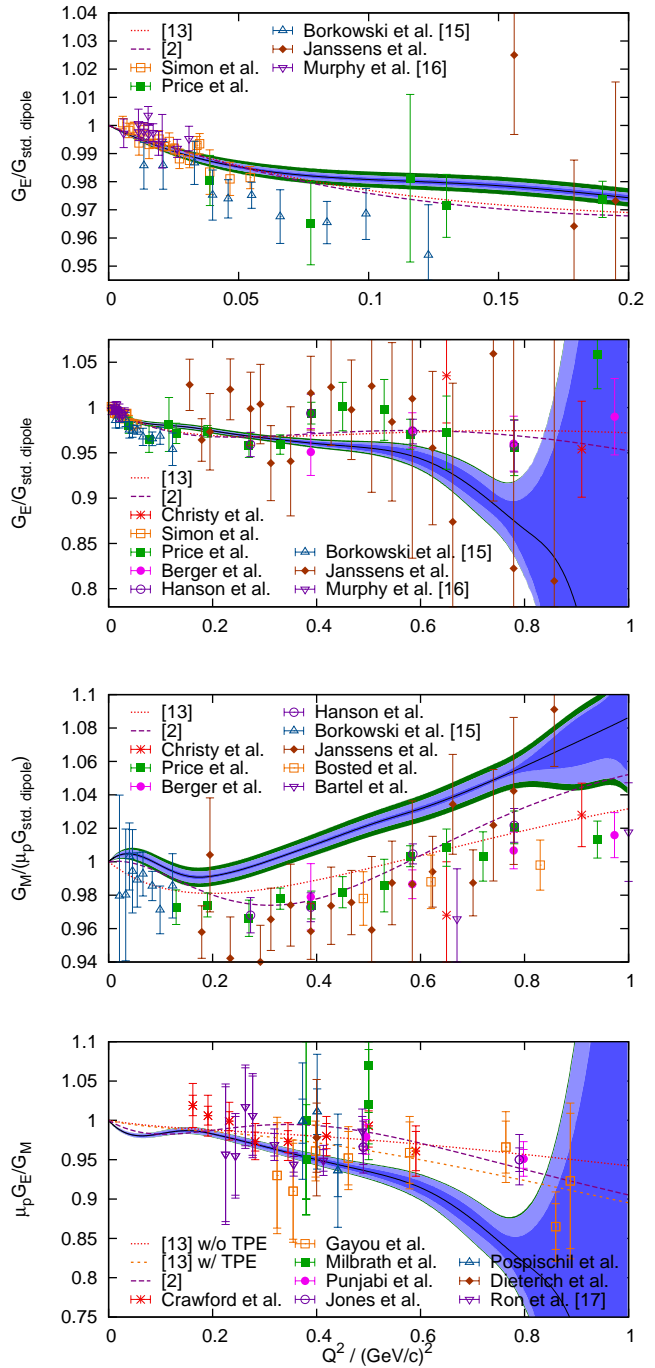} 
\caption{\label{fig_emem} The form factors $G_E$ and $G_M$ normalized 
  to the standard dipole and $G_E/G_M$ as a function of $Q^2$. 
  Black line: best fit to the data, blue area: statistical  
  68\%  pointwise confidence band, light blue area: experimental systematic error,  
  green outer band: variation of the Coulomb correction by $\pm 50\%$. The 
  different data points depict previous measurements, for Refs. see 
  \cite{fw03,Arrington07}; we added the data points of \cite{Borkowski74,Murphy74,Ron07}. Dashed lines are previous fits to the old data in \cite{fw03,Arrington07}.
} 
  \end{figure}  

The form factors extracted with the flexible models agree  among each
other to better than 0.25\% in the $Q^2$ range up to 0.5 $(\mathrm{GeV}/c)^2$. 
They all fit the data equally well with $\chi^2/\mathrm{d.o.f.}\approx 1.14$ 
for $\mathrm{d.o.f.}\approx 1400$.  However, including the less flexible 
models one obtains $1.16 \leqslant \chi^2/\mathrm{d.o.f.} \leqslant 1.29$ and 
the agreement is only better than 0.6\%. In Fig.\,\ref{fig_emem} the results 
of the spline model for $G_E$, $G_M$ and their ratio are shown, together with
previous measurements and fits. The error bars of the previous data shown for 
$G_E$ and $G_M$ are statistical only, normalization uncertainties are typically 
of the order of a few percent. Since TPE corrections are not applied to any of 
the data, the corresponding non-TPE-corrected fit of Ref.\,\cite{Arrington07} is
shown. In the plot of the ratio the fit to the TPE-corrected data of
ref.\,\cite{Arrington07} is also included.
 
The results for $G_E$ exhibit a large negative slope relative to the 
standard dipole at $Q^2 \approx 0$ giving rise to the significantly  
larger charge radius. This slope levels out around 0.1\,$(\mathrm{GeV}/c)^2$ 
and remains constant up to 0.55\,$(\mathrm{GeV}/c)^2$ when the slope again  
becomes larger. In that region, however, only measurements at large  
scattering angles for only two beam energies contribute so that the fit  
becomes less reliable and more sensitive to systematic errors such as  
the neglect of TPE. For even higher $Q^2$ measurements have been taken  
only at one energy and a separation of $G_E$ and $G_M$ is not possible.  
In the close-up for $G_E$ there is an indication of a bump around  
0.15\,$(\mathrm{GeV}/c)^2$, however, at the limit of significance. 
 
The magnetic form factor $G_M$ deviates from earlier measurements. 
This may be related to the normalization at $Q^2 \rightarrow 0$ ignoring 
the wiggle seen by this experiment. The maximum and the minimum 
of the wiggle structure depend, of course, on the parameter of the dipole form.
Also, it is not clear whether the older experiments include the proton 
contribution to the radiative corrections.

The structure at small $Q^2$ seen in both form factors corresponds to 
the  scale of the pion of about  
$Q^2 \approx m_{\pi}^2 \approx 0.02\,(\mathrm{GeV}/c)^2$  
and may be indicative of the influence of the pion cloud \cite{Vanderhaeghen:2010nd}. 
 
While the deviation of $G_M$ from previous measurements seems surprising 
at first glance, it reconciles the form factor ratios from experiments  
with unpolarized electrons, like this one, with those found with polarized  
electrons, especially with the high-precision measurements in  
ref.\,\cite{Ron07}. The previous $G_E$ and $G_M$ data are basically not  
compatible with the polarized measurements even when TPE corrections are 
applied. New results from Jefferson Laboratory \cite{Higinbotham2010,Higinbothampriv}  
with uncertainties of about 2\% confirm this statement and are in excellent 
agreement with this experiment.

The charge and magnetic rms-radii are given by 
\begin{equation} 
\left\langle r^2_{E/M}\right\rangle=-\frac{6\hbar^2}{G_{E/M}\left(0\right)}\left. 
\frac{\mathrm{d}G_{E/M}\left(Q^2\right)}{\mathrm{d}Q^2}\right|_{Q^2=0}. 
\end{equation} 
 
In the study of the model dependency through simulated data only the
flexible models reproduce the input radii reliably. In the fits to the
measured data the models can be divided into two groups: Those based
on splines with varying degree of the basis polynomial and number of
support points and those composed of polynomials with varying
orders. For the charge radius the weighted averages of the two groups
differ by 0.008\,fm.
 
For the spline group we obtain the values 
\begin{eqnarray*} 
\left\langle r_E^2\right\rangle^\frac{1}{2}&=&0.875 (5)_\mathrm{stat.}(4)_\mathrm{syst.}(2)_\mathrm{model}\,\mathrm{fm},\\  
\left\langle r_M^2\right\rangle^\frac{1}{2}&=&0.775(12)_\mathrm{stat.}(9)_\mathrm{syst.}(4)_\mathrm{model}\,\mathrm{fm} 
\end{eqnarray*} 
and for the polynomial group 
\begin{eqnarray*} 
\left\langle r_E^2\right\rangle^\frac{1}{2}&=&0.883(5)_\mathrm{stat.}(5)_\mathrm{syst.}(3)_\mathrm{model}\,\mathrm{fm},\\  
\left\langle r_M^2\right\rangle^\frac{1}{2}&=&0.778(^{+14}_{-15})_\mathrm{stat.}(10)_\mathrm{syst.}(6)_\mathrm{model}\,\mathrm{fm}. 
\end{eqnarray*} 
 
Despite detailed studies the cause of the difference between the two  
model groups could not be found. Therefore, we give as the final  
result the average of the two values with an additional uncertainty  
of half of the difference:
\begin{eqnarray} 
\left\langle r_{E}^2\right\rangle^\frac{1}{2}&=&0.879 (5)_\mathrm{stat.}(4)_\mathrm{syst.}(2)_\mathrm{model}(4)_\mathrm{group}\, 
\mathrm{fm},\nonumber\\  
\left\langle r_{M}^2\right\rangle^\frac{1}{2}&=&0.777 (13)_\mathrm{stat.}(9)_\mathrm{syst.}(5)_\mathrm{model}(2)_\mathrm{group}\, 
\mathrm{fm}.\nonumber 
\end{eqnarray} 
These radii have to be taken with the applied corrections in
mind. While the Coulomb correction used is compatible with other studies  
\cite{Rosenfelder99,Sick03} a more sophisticated theoretical calculation  
may affect the results slightly. 
 
The electric radius is in complete agreement with the CODATA06
\cite{Mohr08} value of $0.8768 (69)$\,fm based mostly on atomic
measurements. It is also in complete accord with the old Mainz result
\citep{Simon80} when the Coulomb corrections
\cite{Rosenfelder99,Sick03} are applied. However, the results from
very recent Lamb shift measurements on muonic hydrogen \cite{pohl} are
0.04\,fm smaller, i.e.\, 5 standard deviations. This difference is
unexplained yet. The calculation of the Lamb shift in muonic hydrogen 
requires the solution of a relativistic bound state problem 
(see Ref.\,\cite{Borie:2004fv} and references therein). The deviation  
may be due to the distorted wave functions, significantly more distorted than in 
electronic hydrogen, necessitating the consideration of multiphoton exchange.

The magnetic radius has a larger error than the charge radius since
the experiment is less sensitive to $G_M$ at low $Q^2$. Its value is
smaller than results of previous fits, however, it is in good agreement
with ref.\,\citep{volotka}, who found $0.778(29)$\,fm from hyperfine
splitting in hydrogen.

The consequences of the results presented here for our picture of the proton are discussed in ref.\,\cite{Vanderhaeghen:2010nd}. A full account of this work will be published \cite{bernauerphd,Bernauer2010:cd}.

\acknowledgments{This work was supported by the Collaborative Research 
Center SFB 443 of the Deutsche Forschungsgemeinschaft. H.\, Fonvieille
is supported by the French CNRS/IN2P3.
}

\bibliography{prl_ff} 
 
\end{document}